\documentclass[journal]{IEEEtran}
\usepackage{graphicx}
\usepackage{subfigure}
\usepackage{float}
\usepackage{amsmath}
\usepackage{amsfonts,amssymb}
\usepackage{dsfont}
\usepackage{algpseudocode}
\usepackage{makecell}
\usepackage{cite}
\usepackage{multirow}
\usepackage{multicol}
\usepackage{booktabs}
\usepackage[ruled,vlined]{algorithm2e}
\usepackage{xcolor,soul}
\usepackage{comment}

\usepackage{fancyhdr,lipsum}

\fancypagestyle{mahmood}{%
   \fancyhf{} 
   
   \fancyhead[C]{You can use this material personally. Reprinting or republishing this material for the purpose of advertising or promotion, creating new collective works, reselling or redistributing to servers or lists, or using any copyrighted component in other works must adhere to IEEE policy. The DOI will be supplied as soon as it becomes available.}
}%

\makeatletter
\let\ps@IEEEtitlepagestyle\ps@mahmood
\makeatother

\begin{document}
\title{Semantic Revolution from Communications to Orchestration for 6G: Challenges, Enablers, and Research Directions}

\author{
    \IEEEauthorblockN{
        Masoud Shokrnezhad\textsuperscript{1}, Hamidreza Mazandarani\textsuperscript{2}, Tarik Taleb\textsuperscript{1, 2}, Jaeseung Song\textsuperscript{3}, and Richard Li\textsuperscript{4}
    }
    \IEEEauthorblockA{
       \\
        \textsuperscript{1} \textit{Oulu University, Oulu, Finland}; \{masoud.shokrnezhad, tarik.taleb\}@oulu.fi \\
        \textsuperscript{2} \textit{Ruhr University Bochum, Bochum, Germany}; hr.mazandarani@ieee.org, tarik.taleb@rub.de \\
        \textsuperscript{3} \textit{Sejong University, Seoul, Korea}; jssong@sejong.ac.kr \\
        \textsuperscript{4} \textit{Futurewei Technologies, USA}; richard.li@futurewei.com
    }
}




\maketitle

\begin{abstract}
In the context of emerging 6G services, the realization of everything-to-everything interactions involving a myriad of physical and digital entities presents a crucial challenge. This challenge is exacerbated by resource scarcity in communication infrastructures, necessitating innovative solutions for effective service implementation. Exploring the potential of Semantic Communications (SemCom) to enhance point-to-point physical layer efficiency shows great promise in addressing this challenge. However, achieving efficient SemCom requires overcoming the significant hurdle of knowledge sharing between semantic decoders and encoders, particularly in the dynamic and non-stationary environment with stringent end-to-end quality requirements. To bridge this gap in existing literature, this paper introduces the Knowledge Base Management And Orchestration (KB-MANO) framework. Rooted in the concepts of Computing-Network Convergence (CNC) and lifelong learning, KB-MANO is crafted for the allocation of network and computing resources dedicated to updating and redistributing KBs across the system. The primary objective is to minimize the impact of knowledge management activities on actual service provisioning. A proof-of-concept is proposed to showcase the integration of KB-MANO with resource allocation in radio access networks. Finally, the paper offers insights into future research directions, emphasizing the transformative potential of semantic-oriented communication systems in the realm of 6G technology.
\end{abstract}


\IEEEpeerreviewmaketitle

\section{Introduction}\label{s_intro}
\begin{figure}[!t]
\centerline{\includegraphics[width=2.5in]{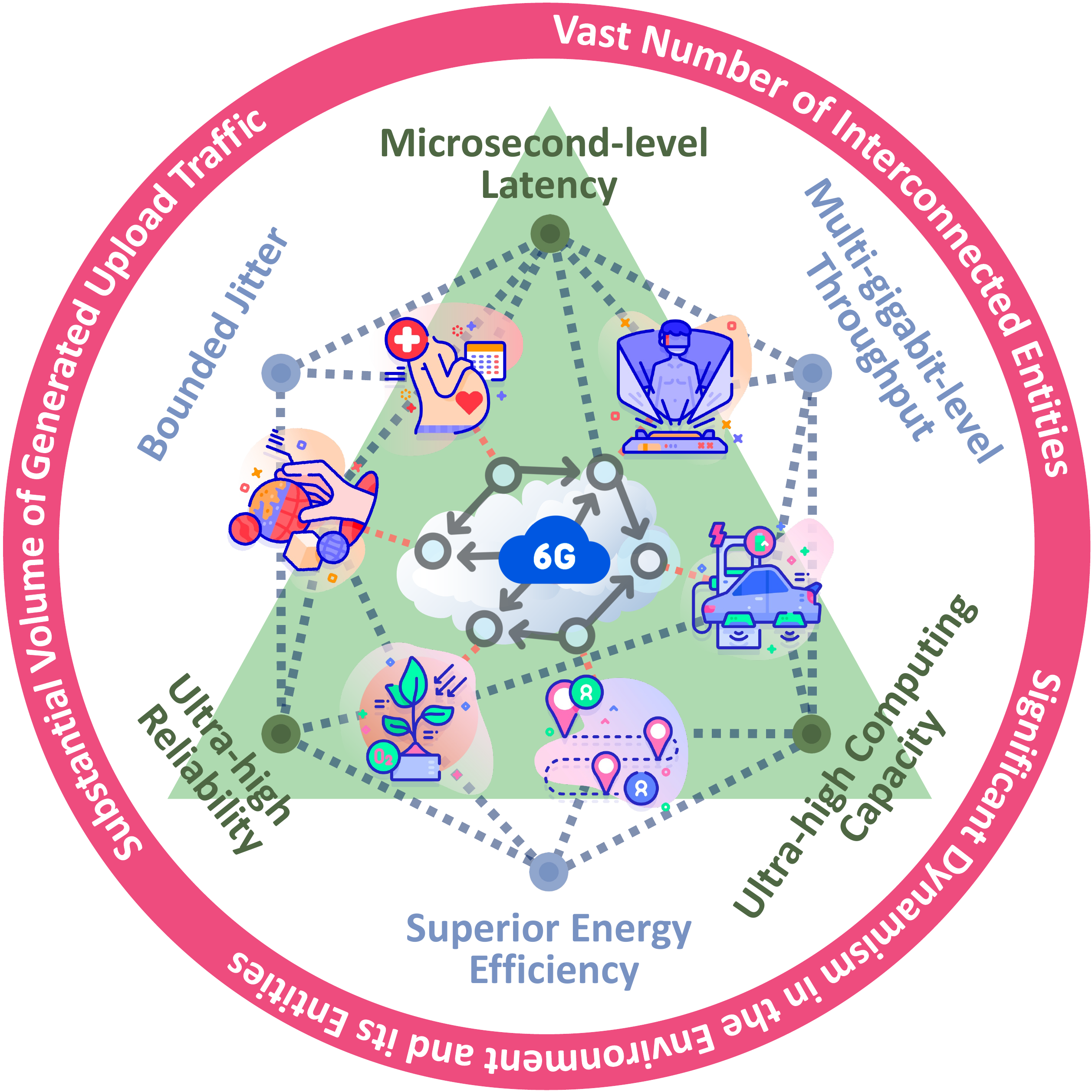}}
\caption{The anticipated characteristics, QoS/QoE criteria, and illustrative examples of services expected in 6G.}
\label{fig1}
\end{figure}

In the foreseeable future of 6G communication systems, connections are expected to expand to everything-to-everything interactions within platforms like the Metaverse, involving diverse physical and digital objects continually engaging, traversing, and coexisting. A consequential outcome of deploying such immersive environments is the continual proliferation of connected objects, leading to significant increases in upstream traffic. Predictions suggest that 6G will be linked to highly dense environments with a considerable number of entities and a substantial volume of global data, of which a significant portion will be routed to computing resources, particularly for services such as telemedicine, holographic teleportation, immersive learning, precision agriculture, smart supply chaining, and intelligent transportation. These services necessitate rigorous End-to-End (E2E) Quality of Service/Experience (QoS/QoE) prerequisites, encompassing requirements such as microsecond-level latency, bounded jitter, multi-gigabit-level throughput, ultra-high reliability, exceptionally high computing capacity, and superior energy efficiency~\cite{9163104}, as illustrated in Fig.~\ref{fig1}.

\begin{figure}[!t]
\centerline{\includegraphics[width=3.5in]{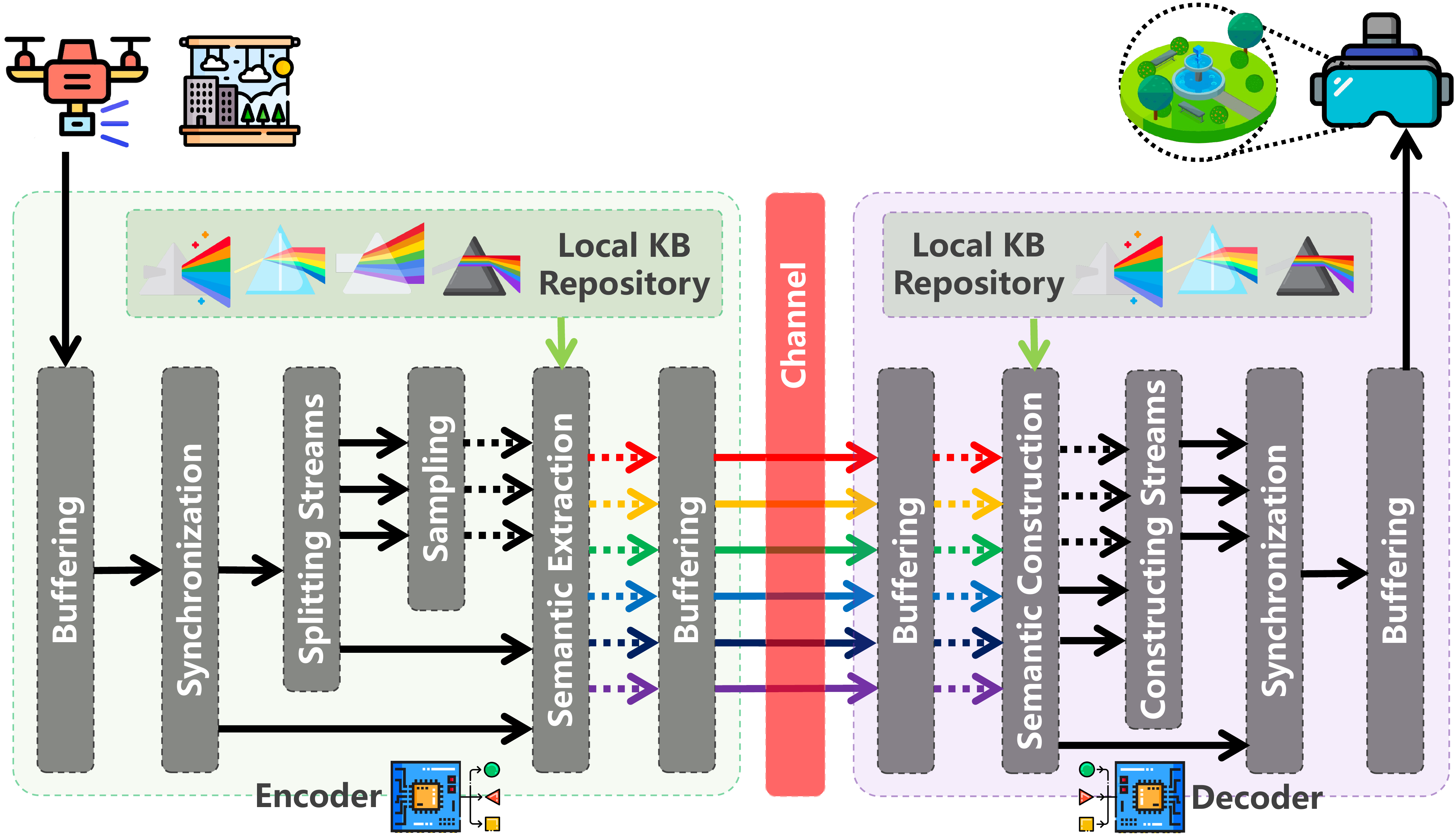}}
\caption{The elements of semantic encoders and decoders (transcoders) encompassing various tasks, including buffering traffic, synchronizing traffic from diverse sources, splitting or constructing streams, sampling data from distinct streams, extracting or constructing semantics, and performing traffic shaping. Semantic encoders can be implemented as service-specific functional components on user devices or the system's edge computing resources. Conversely, semantic decoders should be implemented at locations where the semantics need to be utilized.}
\label{fig2}
\end{figure}

To meet the substantial demand with stringent performance requirements, the scarcity of infrastructure resources emerges as a significant challenge. To tackle this challenge, there has been notable interest in a recent paradigm, recognized as \textit{Semantic Communications} (\textit{SemCom}). This paradigm extends beyond the conventional Shannon paradigm (which primarily optimizes opaque data pipes aiming to reproduce exactly exchanged sequences of symbols) and centers on \textit{effectively} inferring the \textit{meaning} of what has been communicated. This process is facilitated through the utilization of background and contextual knowledge, known as Knowledge Bases (KBs), shared a priori between communicating parties necessary for achieving a predefined shared view on the goal of communication~\cite{sana2022learning}. To facilitate this, the sender and receiver engage in semantic encoding/decoding (transcoding), as detailed in Fig.~\ref{fig2}. This semantic shift promises a ubiquitous connection without mandating the presence of a shared syntax or protocol in the transmission of data, improves communication efficiency and reliability, and enhances QoS/QoE for human-oriented services~\cite{9530497}. 
Furthermore, in contrast to conventional bit-oriented methodologies where there is no inherent meaning beyond bits, semantics can play a pivotal role in the decision-making process of orchestrating communication system resources. Leveraging the correlation between semantics allows for the optimization of resource scheduling and utilization.

The exploration of realizing SemCom and its impact on communication efficiency has been a focal point in recent research efforts. Some studies have investigated the potential of semantic-oriented communications to enhance point-to-point physical layer connections, aiming to surpass Shannon's limit. Meanwhile, new research explores incorporating semantic awareness into emerging communication systems. Trevlakis \textit{et al.}~\cite{10454584} thoroughly examined the integration of SemCom in the context of 6G, suggesting a new SemCom network structure based on optimal point-to-point resource utilization. Luo \textit{et al.}~\cite{9679803} and Wheeler \textit{et al.}~\cite{Wheeler2022Engineering} proposed ML-aided E2E semantic-aware communication systems for beyond 5G. Yang \textit{et al.}~\cite{9979702} introduced a similar edge-driven concept with the aim of improving transcoding efficiency, where training, maintenance, and execution of semantic transcoding are based on KBs shared by service users. Similarly, Lu \textit{et al.}~\cite{lu2023semantics} developed an ML-transcoded semantic-aware design inspired by natural human interactions. Furthermore, some research efforts have aimed to introduce semantic awareness into radio network architecture through architectural enhancements~\cite{Iyer2022A, li2023open, strinati2024goal}.


While the aforementioned studies have tried to address different challenges to realize SemCom, their applicability to 6G is limited. Achieving efficient SemCom in 6G necessitates universal access to the same background knowledge by all semantic transcoders of a given service, posing an exceptionally formidable challenge, especially for dynamic and non-stationary future systems \cite{9163104, 9679803, Iyer2022A, lu2023semantics}. Existing literature lacks a real-time and systematic approach to address the challenge of KB management, particularly tailored to the structure of 6G and its ever-fluctuating services. To address this gap, we first describe future 6G systems and their components in Section \ref{s_bck}, highlighting the most challenging aspects of enabling SemCom: KB refinement and KB arrangement. Subsequently, we introduce the KB Management And Orchestration (KB-MANO) framework in Section \ref{s_method}, designed for the allocation of network and computing resources dedicated to updating and redistributing KBs across the system. The primary objective is to minimize the impact of knowledge management on actual service provisioning. Given that KB-MANO is grounded in the principles of Computing-Network Convergence (CNC) and lifelong learning, this section also delves into strategies suitable for KB updating across various scenarios. Section \ref{s_eva} features a proof-of-concept scenario evaluating the efficiency of semantic-aware orchestration enabled by KB-MANO, followed by an exploration of open research directions in Section \ref{s_frd}. The paper concludes in Section \ref{s_con}, summarizing key findings and implications.

\section{Fundamentals, Challenges, \& Enablers} \label{s_bck}

\subsection{Service Provision over Integrated 6G}
\begin{figure*}[!t]
\centerline{\includegraphics[width=7.2in]{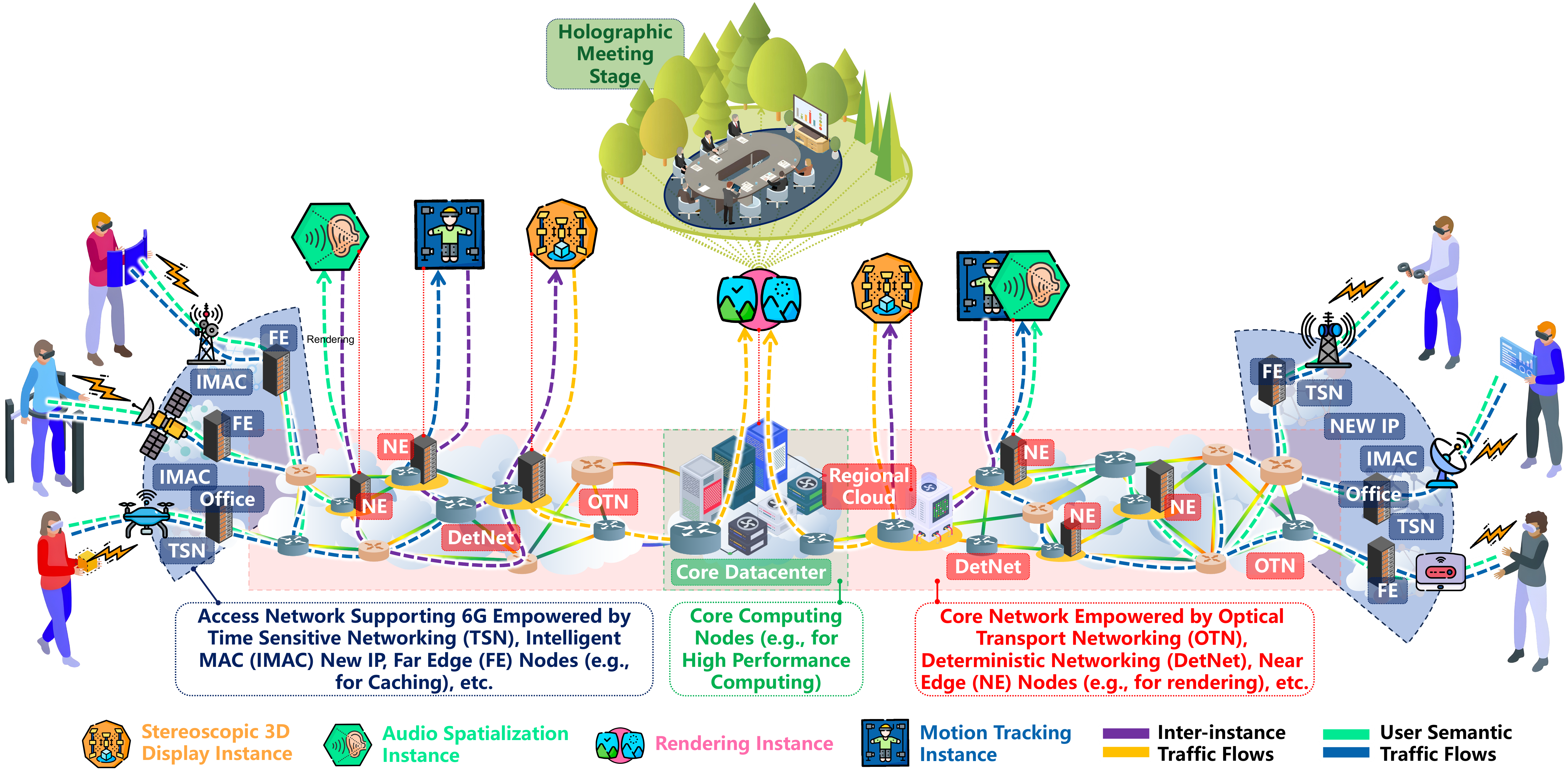}}
\caption{Forthcoming 6G services facilitated by an integrated cloud-network infrastructure, incorporating technologies such as deterministic networking, time-sensitive networking, and intelligent medium access control.}
\label{fig3}
\end{figure*}

In the context of 6G systems, decentralized computing resources extend across in-network, edge, regional, and central nodes dispersed across vast geographical areas. These nodes interconnect through diverse networking technologies in radio access, transmission, and core network sub-domains, featuring varied network devices and links. Provisioning future services over such infrastructures necessitates precise resource orchestration. A Metaverse scenario depicting users participating in a holographic presence service within a virtual conference room by a live lakeshore is illustrated in Fig.~\ref{fig3}. The service integrates rendering, motion tracking, stereoscopic 3D display, and audio spatialization functionalities. To bring this experience to life, the software instances of these functionalities should be loaded onto available computing nodes. Subsequently, users' video, audio, and motion data are transmitted to the instances through their designated network paths. These instances collaborate in accordance with a predetermined order specified in the service's function chaining map. Following that, the generated rendered content is transmitted back to users' devices, such as headsets. Throughout these processes, strict compliance is maintained with factors such as the QoS/QoE requirements of the service (as delineated in Fig.~\ref{fig1}) and the availability of resources.

\subsection{Semantic Mastery in Orchestration}
In the context of SemCom for future extensive-scale services, users employ semantic encoders to compress data into semantic segments (or semantics), transmitting these semantics instead of raw bulk data to their designated instances. The instances, in turn, utilize semantic decoders to reconstruct the intended output. In scenarios such as holographic conferences, spatial, visual, and auditory semantics—including positioning, gestures, and ambient sounds—are extracted using semantic encoders, and semantic decoding captures user interactions with virtual elements, each other, and the environment, contributing to a realistic rendering and presentation of the holographic presence. The adoption of semantics not only reduces resource consumption and enhances efficiency but also optimizes resource orchestration. By understanding the semantics of user requests, prioritization can be applied giving top priority to critical semantics to ensure seamless service responsiveness, and similar requests can be directed to shared computing nodes and network paths. While the vision of SemCom has long existed, the recent emergence of ML-aided semantic understanding is recognized as a promising enabler of 6G \cite{getu2024performance}. This involves the utilization of semantic-specific Deep Neural Networks (DNNs) for transcoding semantics\footnote{In ML-aided SemCom, background knowledge corresponds to the DNN weights of a trained ML agent. In non-ML solutions, it may take the form of a knowledge graph or a set of rules extracting information bit values.}.

\subsection{Semantic Costs in Play}
In achieving effective semantic-aware provision and orchestration, a significant challenge arises from the dynamic nature inherent in future systems. To comprehend this dynamism, consider the holographic meeting scenario, wherein the ever-changing nature of interactions is influenced by a multitude of sources. For instance, users may fluidly transition between various roles, each accompanied by distinct QoS requirements, or they may modify their receivers, evolving their QoE standards for immersive experiences across each receiver. Additionally, physical mobility at varying speeds introduces changes in users' connection points to the system. Considering the temporal fluctuations in 6G resources, distributed across diverse domains and operating under different supervisions, two primary challenges are anticipated: \textit{knowledge refinement} and \textit{knowledge arrangement}. Knowledge refinement pertains to the optimal allocation of scarce computing and network resources to continuously evolve KBs, and knowledge arrangement involves the efficient distribution of updated KBs to users, instances, and entities responsible for resource allocation, such as network edge devices and/or the system orchestrator. In dynamic environments, constantly refining KBs while arranging every piece of updated knowledge can lead to impractical resource depletion. Thus, achieving a balanced approach is essential for optimal semantic-aware provisioning and orchestration.

\subsection{The Era of Intelligent Integration}
Addressing knowledge refinement involves utilizing the concept of \textit{lifelong learning}, which accommodates a theoretically unlimited number and variety of tasks, segregating training procedures for different contexts by assigning each one to a task. This results in the long-term applicability and coherence of trained tasks for recurring scenarios, diminishing the need for frequent retraining, as well as the effectiveness of training from scratch for new situations. The approach mitigates the environmental impact and computational overhead associated with traditional ML techniques. In the realm of SemCom, this proves advantageous for adeptly managing emerging situations. For instance, in the holographic meeting scenario, lifelong learning can be applied to 1) dedicate an individual task to maintain each existing KB updated with changes in data streams (such as emerging words and meanings in supported languages) and 2) initiate new tasks in response to the demand for new KBs (such as the introduction of new data streams facilitated by emerging user-side technologies).

Another crucial enabler is the concept of CNC. In the context of the integrated infrastructure of 6G, consisting of distinct computing and network domains, CNC aims to orchestrate diverse domain resources collectively, establishing network-aware, detectable, assignable, and schedulable computing resource pools. By actively monitoring computing and network domains in real-time and discerning system-wide states, the CNC E2E orchestrator adeptly allocates resources for each domain, facilitated not only by leveraging its own domain state but also by incorporating the state information from the other domain. This dynamic adjustment considers changes in network accessibility, availability, and computing capacity quality \cite{shokrnezhad2024acnc}, stemming from factors like dynamic traffic patterns or administrative restrictions imposed by different domains or regions. Employing the CNC concept ensures that data collection, the training process for KBs, and subsequent distribution among users, instances, and entities making resource allocation decisions are conducted with careful consideration to prevent overloads on actual network traffic and running services.

\section{Proposed Framework} \label{s_method}
\subsection{The System Model}
To enable semantic-aware provision and orchestration for 6G's dynamic, massive-scale services with stringent QoS/QoE requirements, we consider a system equipped with integrated computing and network resources, referred to as \textit{the infrastructure}. Within this system, \textit{service providers} (like those facilitating the holographic meeting scenario) register services, encompassing functional instances (e.g., rendering, motion tracking, etc.). The registered services incorporate SemCom, enabled by their corresponding KBs, and introduce their own lifelong learning tasks (or simply, training tasks) that ensure KBs remain updated. \textit{Users} initiate requests to access the registered services within the system, establishing E2E connections that manage the transmission of semantics to the instances of their requested services. Processed data (such as a live-rendered holographic meeting scene) is then delivered back to users. Resource allocation for service instances and user requests occurs through the multi-layer, CNC-empowered resource orchestration framework, named CNCO, implementing decisions at the network's edge devices, named the Points of Arrival (PoAs), the entry points for requests into the system. CNCO employs KBs associated with each service at PoAs to comprehend the transmitted semantics, integrating them into its decision-making process.

\begin{figure*}[!t]
\centerline{\includegraphics[width=7.1in]{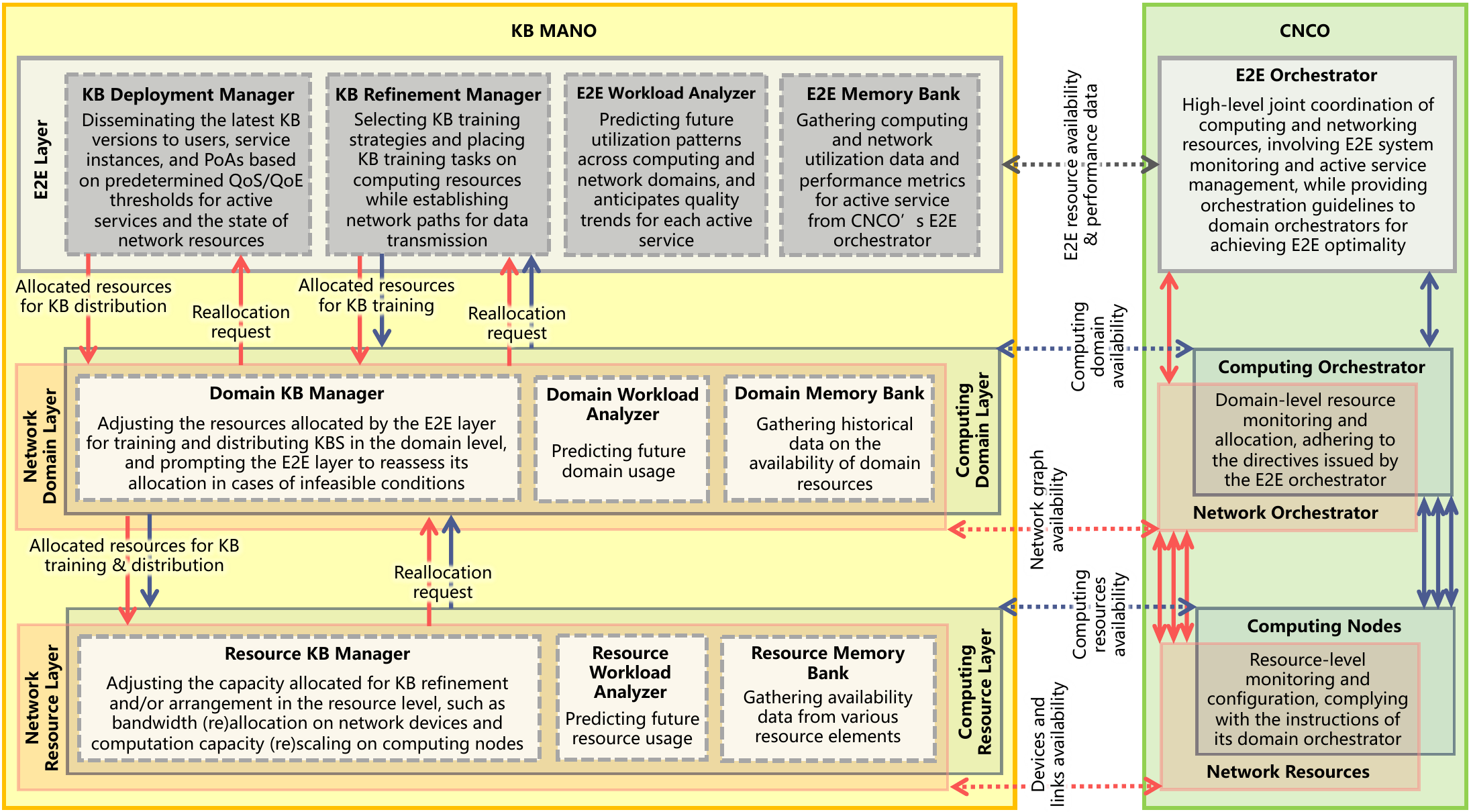}}
\caption{The KB-MANO architecture integrated with the CNCO framework (detailed by Shokrnezhad \textit{et al.} \cite{shokrnezhad2024acnc}). It is noteworthy that the E2E and domain layers can be instantiated as components within the CNCO E2E and domain orchestrators, respectively. Furthermore, the network and computing resource layers can be realized as virtual functions on network devices and computing nodes.}
\label{fig4}
\end{figure*}

\subsection{KB-MANO}
Considering the system model and adhering to the defined challenges of knowledge refinement and arrangement, we intend to optimize the allocation of network and computing resources for updating and (re)distributing KBs throughout the system. This ensures that users, service instances, and PoAs have access to the latest KBs, aiming to minimize the impact of knowledge management and orchestration activities on actual service provisioning. To achieve this objective, we propose the KB-MANO framework, which addresses knowledge refinement by incorporating KB training strategies enabled by lifelong learning and dynamically switching between them based on available resource information received from CNCO. Additionally, it handles knowledge arrangement by managing the organization of KB distribution among users, instances, and PoAs through proactive measures initiated by service-level QoS/QoE degradation thresholds, guided by monitoring information provided by CNCO. The components of KB-MANO are categorized into three distinct layers: \textit{E2E}, \textit{Domain}, and \textit{Resource}, as depicted in Fig.~\ref{fig4} and elaborated upon in the following subsections.

\vspace{10px} \subsubsection{E2E layer}\label{sss_e2e} \hfill \vspace{3px} \\ \indent 
The \textit{E2E} layer assumes the responsibility of high-level decision-making related to knowledge refinement and arrangement. Its initial function involves gathering computing and network utilization data from CNCO's E2E orchestrator, which is then stored in the \textit{Memory Bank} over time. Additionally, performance metrics for active services across all users are received from CNCO. Note that the collection of this information is conducted anonymously to uphold user privacy. The \textit{Workload Analyzer} component accumulates this historical data over the last $\mathcal{T}^e$ time slots, predicts future utilization patterns across computing and network domains, and anticipates QoS/QoE trends for each active service. The derived insights are subsequently stored in the memory bank. Evidently, $\mathcal{T}^e$ (the history window size) is adjustable, catering to the dynamics of the system. Smaller window sizes capture recent fluctuations with computational efficiency, while larger window sizes yield more stable predictions using more complex prediction models. ML techniques, such as eXplainable Long Short-Term Memory (XLSTM) and transformers, are employed to analyze historical and temporal data, enabling the forecasting of future states.

Now, the \textit{KB Refinement Manager} relies on present and anticipated resource states to allocate resources for KB updates. This entails two primary steps: selecting a KB training strategy, which assesses the suitability of distributed versus centralized processing based on current and future resource availability (elaborated in Section \ref{ss_str}), and subsequently placing the KB training tasks on computing resources while establishing network paths for data transmission from users to the assigned computing nodes. As these training tasks are implemented and executed by the service provider (for example, the entity in charge of the holographic meeting setup), the data remains in their custody, alleviating privacy concerns. The conclusive module, designated as the \textit{KB Deployment Manager}, is tasked with disseminating the latest KB versions to users, service instances, and PoAs. The distribution process hinges on predetermined QoS/QoE thresholds for active services and the current and anticipated network resources. A heightened sensitivity to slight quality alterations prompts increased network resource utilization for KB distribution, ensuring continual updates across the system. Conversely, adopting more lenient thresholds leads to reduced network consumption, albeit potentially resulting in less precise semantic transcoding.

\vspace{10px} \subsubsection{Domain Layer}\label{sss_dom} \hfill \vspace{3px} \\ \indent 
The \textit{Domain} layer oversees intra-domain resource allocation for knowledge refinement and arrangement. In the network domain, historical data on the availability of the network graph is received from CNCO's network orchestrator and stored in the \textit{Memory Bank}, along with usage predictions from the domain-level \textit{Workload Analyzer}. The history window size in this layer ($\mathcal{T}^d$) can be set smaller than $\mathcal{T}^e$ to capture transient changes with reduced complexity. Based on the strategy chosen by the \textit{KB Refinement Manager} and current and predicted network usage, the \textit{Domain KB Manager} adjusts network paths for data transmission (between users and the allocated computing nodes) and KB distribution. In the absence of a feasible network path for either conveying user data or distributing the latest KBs, this module prompts the \textit{E2E} layer to reassess its chosen strategy, optimizing it to align with the current network conditions. Path feasibility is compromised when the necessary bandwidth surpasses the thresholds established to protect actual network traffic. Another factor is the excessive latency of bandwidth-feasible paths, preventing the timely update of KBs within the predefined time constraints dictated by the QoS/QoE requirements of their associated services. The computing domain mirrors this process, readjusting computing capacity for assigned tasks.

\begin{figure*}[!t]
\centerline{\includegraphics[width=7.2in]{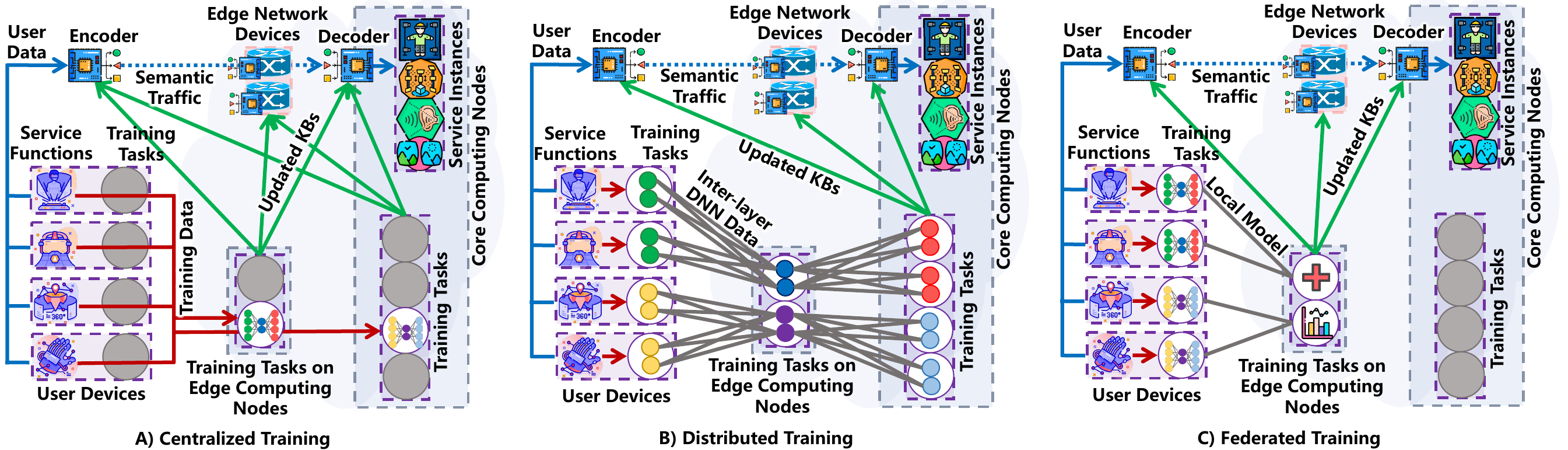}}
\caption{The KB refinement training strategies, including centralized, distributed, and federated approaches. It is essential to highlight that the implementation of the semantic encoder can occur independently on user devices or be integrated into network edge devices.}
\label{fig5}
\end{figure*}

\vspace{10px} \subsubsection{Resource Layer}\label{sss_res} \hfill \vspace{3px} \\ \indent 
The \textit{Resource} layer manages resource allocation for knowledge refinement and arrangement across network devices and computing nodes. It gathers availability information from various resource elements, predicts their future states, and stores this data in the resource \textit{Memory Bank}. The history window size in this layer ($\mathcal{T}^r$) is kept minimal to reduce resource consumption during the prediction process and capture subtle resource-level fluctuations. Subsequently, the \textit{Resource KB Manager} utilizes this information to dynamically adjust the capacity allocated for KB refinement and/or arrangement. On network devices, this adjustment includes (re)allocating bandwidth for transmitting user data from users to training tasks or prioritizing traffic for distributing the latest KBs from training tasks to users, service instances, or PoAs. On computing nodes, the adjustment entails (re)scaling allocated computation, memory, and storage capacity for training tasks or migrating them to nodes accessible via the assigned network paths. In cases of infeasibility, the \textit{Domain} layer is notified by the \textit{KB Refinement Resource Manager} to reallocate resources at the domain level.

\subsection{KB Training Strategies}\label{ss_str}
Concerning the system state, which denotes the current availability and the predicted availability of resources for future time slots, the \textit{KB Refinement Manager} in the \textit{E2E} layer can adopt various strategies for updating KBs. As illustrated in Fig.~\ref{fig5}, the most favorable strategies include:

\vspace{10px} \subsubsection{Centralized Training}\label{sss_str1} \hfill \vspace{3px} \\ \indent  
The straightforward strategy is to treat each KB's training tasks as an atomic operation and place them on computing nodes, either at the network edge or in core data centers, to be executed in a centralized manner as illustrated in Fig.~\ref{fig5}-A. This type of offloading simplifies the process by avoiding the complexities of task decomposition and combinatorial resource optimization, which could introduce additional computation costs and scheduling delays in the \textit{E2E} layer. However, this strategy requires substantial data transmission from users to edge or core computing nodes, which may not be viable in future scenarios where limited network resources must be allocated to actual traffic. Additionally, due to the atomic nature of training task placement, this strategy lacks flexibility for accommodating future dynamic patterns and may lead to inefficient resource utilization. As another notable vulnerability, the transmission of user data to edge or core computing nodes may potentially elevate the risk of privacy concerns.

\vspace{10px} \subsubsection{Distributed Training}\label{sss_str2} \hfill \vspace{3px} \\ \indent 
Considering the enhanced capabilities of future user devices, a viable strategy involves partially offloading training tasks to edge or core computing nodes, as depicted in Fig.~\ref{fig5}-B. In this strategy, the \textit{E2E} layer decomposes each KB's training tasks into partitions, and it determines the allocation of these partitions to user devices and the infrastructure's computing nodes. While the dynamic adjustment of partitioning and resource allocation poses a complex optimization challenge, it enables more efficient resource utilization by considering the real-time and predicted states of resources. A potential decomposition method within this strategy involves breaking tasks at the DNN level, where an individual or groups of DNN layers form one partition (a.k.a. split learning). Executing these partitions across user devices and computing nodes in a distributed manner achieves the execution of the entire DNN. The classification of partitions as computation- or network-intensive is then determined based on the live and predicted resource states, guiding their placement on user devices or computing nodes.

\vspace{10px} \subsubsection{Federated Training}\label{sss_str3} \hfill \vspace{3px} \\ \indent 
To significantly mitigate privacy and security risks, as well as reduce the volume of data transmission between users and computing nodes during KB updates, training tasks can be performed on user devices using their local data. However, since the data of an individual device may not adequately represent global coherence, the resulting trained models (e.g., updated DNN weights) can be transmitted to edge and core computing nodes for hierarchical aggregation (such as averaging DNN weights). Guided by the live and predicted states of resources and the characteristics of services, including their QoS/QoE metrics, the \textit{KB Deployment Manager} intermittently replaces local models with global models. This transition ensures that the training process can resume over the updated models. The closed-loop nature of this strategy is illustrated in Fig.~\ref{fig5}-C. While this strategy is well-suited for potential offline scenarios, it does elevate the load on user devices, potentially leading to increased energy consumption and associated costs for users.

\section{Performance Evaluation} \label{s_eva}
In this section, we present a proof-of-concept scenario designed to assess the effectiveness of the KB-MANO framework in facilitating semantic-aware orchestration. Specifically, we investigate a radio cell served by a Small Base Station (SBS), where $\mathcal{N}$ intelligent users contend for access to $\mathcal{C}$ time-slotted uplink channels allocated to the SBS. Collisions occur when multiple users attempt to transmit data over the same channel within the same time slot. The aim is to illustrate that by enabling users to extract semantic information from their transmitted data and subsequently sharing this knowledge with the SBS through the implementation of KB-MANO, the semantic throughput (or simply throughput), defined as the number of successful semantic transmissions, can be improved. To accomplish this, we utilize a Double and Dueling Deep Q-Learning (D3QL)-based approach to categorize users' data into a predefined set of $\mathcal{K}$ semantics. The training and execution of this model, as well as the sharing of its weights (KBs) with the SBS, are facilitated through the application of federated training, as discussed in Section~\ref{sss_str3}.

Subsequent to the extraction of semantic information by the SBS, we employ a method termed Semantic Aware Multiple Access (SAMA)-D3QL, serving as part of OCNC as detailed in \cite{our_cl_d3ql_paper}, to manage user channel access. Throughout the training phase of this approach, each user constructs a historical record, wherein each entry pertains to a particular time slot, encompassing:
\begin{itemize}
    \item User's action:
    \begin{itemize}
        \item $(Sen., c)$: Sense channel $c$.
        \item $(Trn., c)$: Transmit over channel $c$.
    \end{itemize}
    \item User's observation:
    \begin{itemize}
        \item When sensing: $\{Busy, Idle\}$.
        \item After transmissions: $\{Success, Collision\}$.
    \end{itemize}
    \item User's assisted throughput: The number of shared semantics between this user and others, transmitted by this user during each time slot. The SBS calculates this metric utilizing the transmission records of users, enabled by the deployment of KB-MANO.
\end{itemize}
By configuring the reward as the weighted average of throughputs, with the weights corresponding to assisted throughputs, each user trains its individual SAMA-D3QL model. Subsequently, users employ their respective models to make determinations regarding medium access.

\begin{figure}[!t]
\centerline{\includegraphics[width=3.3in]{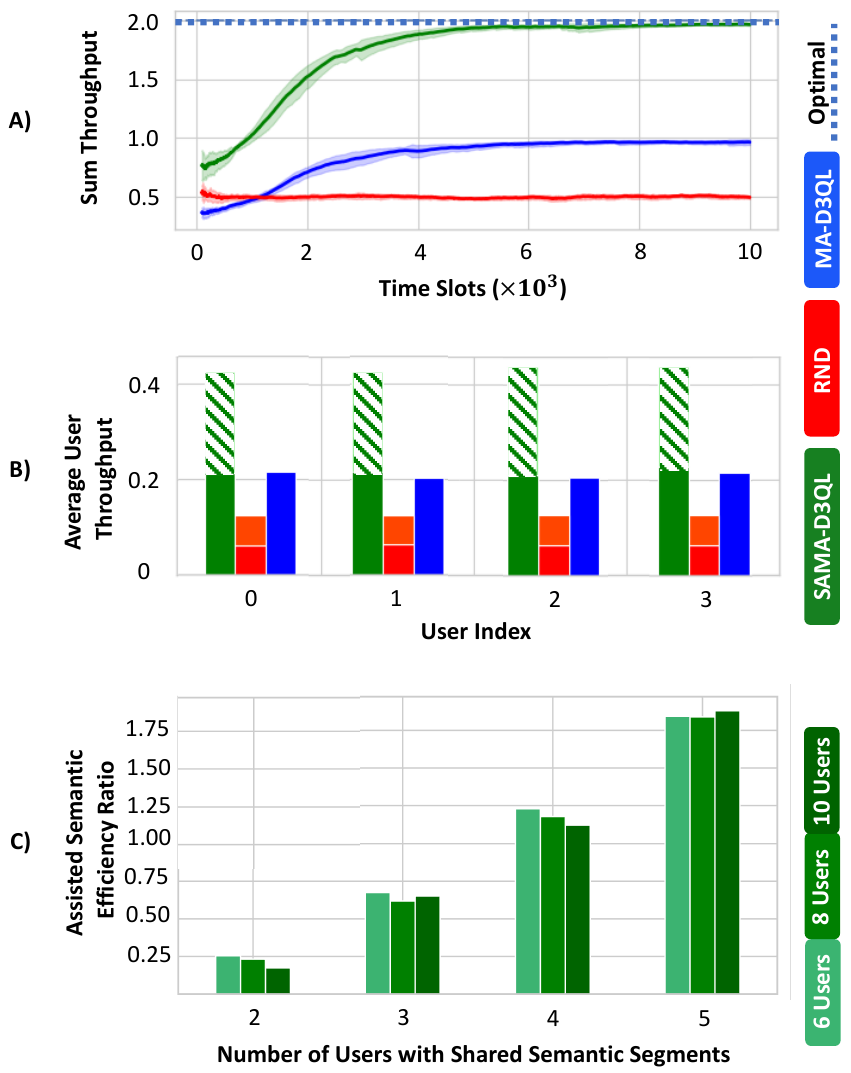}}
\caption{A) The evolution of D3QL-based methodologies throughout their operational time slots, compared with the random and optimal outcomes, in terms of sum throughput, and B) the average throughput for all users and methodologies. It is pertinent to note that, in the experimental setup of A and B, $\mathcal{N}$ is set to 4, $\mathcal{C}$ is 1, and users are divided into two distinct groups, each sharing similar data and corresponding semantics among their group members. C) The assisted semantic efficiency ratio of SAMA-D3QL, computed as the average of assisted throughput divided by total throughput for all users, for varying numbers of all users and different quantities of users with shared semantic segments.}
\label{fig6}
\end{figure}

In Fig.~\ref{fig6}-A and -B, we compare the outcomes of SAMA-D3QL against those of MA-D3QL (without assisted throughput data), random access control (RND), and the optimal solution derived from exhaustive search. Fig.~\ref{fig6}-A illustrates the temporal evolution of total throughput, revealing SAMA-D3QL's significant outperformance of MA-D3QL, ultimately converging to the optimal solution. Fig.~\ref{fig6}-B portrays average user throughputs, with the shaded region denoting each user's assisted throughput, underscoring the spectrum utilization enhancement achieved through semantic awareness. In Fig.~\ref{fig6}-C, the assisted semantic efficiency ratio of SAMA-D3QL, computed as the average of assisted throughput divided by total throughput for all users, is depicted for varying numbers of users and different quantities of users with shared semantics. It is observed that the increase in the number of users sharing semantics results in enhanced assisted throughput. Specifically, with 5 users sharing semantics, each transmission assists $\sim$1.75 semantics. This indicates that with KB-MANO implementation,  $\sim$1.75 transmissions can be omitted for 1 successful transmission, resulting in significant resource savings. These freed resources can then be allocated to transmit larger or redundant semantics to achieve higher-level semantic metrics and distribute updated KBs among network elements. Notably, similar resource savings are anticipated for computing resources. Moreover, Fig.~\ref{fig6}-C demonstrates the scalability of KB-MANO implementation in communication infrastructures, as this phenomenon remains consistent regardless of the number of users.

\section{Future Research Directions}\label{s_frd}
To achieve an efficient and effective implementation of semantic-enabled 6G systems, meticulous consideration must be devoted to addressing various technological and societal challenges, some of which are outlined in the following:

\vspace{10px} \subsubsection{E2E Management of KB Refinement and Deployment}\label{sss_emrd} \hfill \vspace{3px} \\ \indent 
Within the \textit{E2E} layer of KB-MANO, two crucial components stand out: the \textit{KB Refinement and Deployment Managers}. These components address optimization problems related to resource allocation, including running KB training tasks, facilitating data transmission between users and these tasks, and distributing KBs across the infrastructure. In scenarios where federated training is enabled (see Section~\ref{ss_str}), training tasks occur on user devices, eliminating the need for data transmission during training, and model weights (which have negligible size compared to user data) are transmitted to distribute KBs. However, for distributed or centralized training (when device resources are insufficient), training tasks must migrate to edge or cloud resources, requiring data transmission between users and these tasks. In these cases, precise allocation of computing and network resources, taking into account service characteristics and real-time system state, is essential to minimize the impact on actual traffic. Challenges include selecting and configuring training strategies, allocating network bandwidth for KB updates and computing power for model trainings, and prioritizing traffic for deploying updated KBs. ML-based or heuristic approaches offer potential solutions within specified time constraints and considering the dynamic nature of the system, representing a promising avenue for future research.

\vspace{10px} \subsubsection{E2E Calculation of Semantic Performance}\label{sss_epm} \hfill \vspace{3px} \\ \indent 
In the E2E management of KB refinement and arrangement, a crucial consideration lies in constraining service characteristics, particularly semantic-based service performance requirements. Unlike traditional networks with well-defined E2E performance parameters, translating the E2E QoS/QoE performance of semantic-enabled systems into metrics such as E2E latency or minimum required network and computing capacity poses challenges. For example, speech-based services commonly utilize metrics like Word Error Rate (WER) and Signal-to-Distortion Ratio (S2DR), while video-based services rely on Video Quality Assessment (VQA) metrics for performance evaluation~\cite{getu2023making}. Furthermore, the subjective nature of semantics adds complexity, as different services, users, or service providers in multi-domain scenarios may perceive performance differently. Addressing semantic-based metrics and transforming them into constraints and thresholds for resource allocation through innovative research initiatives will aid in overcoming current challenges associated with assessing system-wide semantic awareness comprehensively.

\vspace{10px} \subsubsection{E2E Deterministic Provision of Services}\label{sss_egsp} \hfill \vspace{3px} \\ \indent 
In the holistic oversight of E2E KB refinement and arrangement, it is crucial to integrate not solely the service characteristics but also the real-time system state. Among these, a pivotal component is the real-time states of semantic transcoding blocks, in conjunction with the examination of resource states enabled by CNCO. For example, live services like the holographic meeting may involve intensive semantic extraction during certain time slots when users initiate the use of a specific device. In this time frame, considering semantic extraction delays is crucial, which influences E2E quality metrics and makes edge resources the only viable option to implement corresponding service instances. Conversely, during time slots requiring lighter semantic extraction with negligible delay, instances can be migrated to cost-effective regional or core data centers. Thus, dynamic resource allocation, considering the impact of reduced performance due to semantic transcoding, is imperative for E2E deterministic semantic awareness.

\vspace{10px} \subsubsection{Sustainable Design  of Semantic-Aware Systems}\label{sss_sdsas} \hfill \vspace{3px} \\ \indent 
In the pursuit of service providers' strategic objectives, refining and deploying KBs can be strategically structured to enhance various operational metrics. This includes reducing resource utilization, decreasing E2E latency, and augmenting user support capacity. However, integrating SemCom and orchestration requires incorporating semantic transcoding at various stages within the framework, leading to increased demand for computing resources and higher energy consumption. Hence, alongside these objectives, sustainably responsible resource allocation becomes paramount. This involves prioritizing renewable energy sources for computing resources and consolidating tasks and traffic to free up computing and network space, enabling surplus resource deactivation. Moreover, the advent of novel semantic transcoding paradigms, anchored in causality and advanced cognitive capabilities, along with lifelong learning mechanisms such as Continual Learning~(CL) that systematically integrate new information to prevent catastrophic forgetting, presents potential for mitigating unnecessary retrains or redistributions. This advancement holds promise for nurturing more environmentally sustainable, semantically-enabled systems.

\section{Conclusion}\label{s_con}
This paper focused on addressing challenges related to KB refinement and arrangement to enable efficient SemCom for dynamic 6G services. It first introduced the 6G infrastructure and its entities, explained the concept of SemCom, highlighted the benefits of incorporating semantics in resource orchestration, outlined challenges in implementing efficient semantic-enabled 6G systems, and proposed enabling technologies to overcome these challenges. Subsequently, the KB-MANO framework was introduced, designed for allocating network and computing resources dedicated to updating and distributing KBs across the system, aiming to minimize the impact of knowledge management on service provisioning. Additionally, a proof-of-concept scenario demonstrated semantic-aware radio resource orchestration empowered by KB-MANO. The paper concluded by outlining various research avenues that are crucial for the advancement of semantic-oriented communication systems. These directions are essential for paving the way toward an optimized, resilient, inclusive, and future-proof 6G infrastructure.



\bibliographystyle{IEEEtran}
\bibliography{IEEEabrv,main}

\begin{thebibliography}{10}
\providecommand{\url}[1]{#1}
\csname url@samestyle\endcsname
\providecommand{\newblock}{\relax}
\providecommand{\bibinfo}[2]{#2}
\providecommand{\BIBentrySTDinterwordspacing}{\spaceskip=0pt\relax}
\providecommand{\BIBentryALTinterwordstretchfactor}{4}
\providecommand{\BIBentryALTinterwordspacing}{\spaceskip=\fontdimen2\font plus
\BIBentryALTinterwordstretchfactor\fontdimen3\font minus \fontdimen4\font\relax}
\providecommand{\BIBforeignlanguage}[2]{{%
\expandafter\ifx\csname l@#1\endcsname\relax
\typeout{** WARNING: IEEEtran.bst: No hyphenation pattern has been}%
\typeout{** loaded for the language `#1'. Using the pattern for}%
\typeout{** the default language instead.}%
\else
\language=\csname l@#1\endcsname
\fi
#2}}
\providecommand{\BIBdecl}{\relax}
\BIBdecl

\bibitem{9163104}
L.~U. Khan, I.~Yaqoob \emph{et~al.}, ``{6G Wireless Systems: A Vision, Architectural Elements, and Future Directions},'' \emph{IEEE Access}, vol.~8, pp. 147\,029--147\,044, 2020.

\bibitem{sana2022learning}
M.~Sana and E.~C. Strinati, ``{Learning Semantics: An Opportunity for Effective 6G Communications},'' in \emph{2022 IEEE 19th Annual Consumer Communications \& Networking Conference (CCNC)}.\hskip 1em plus 0.5em minus 0.4em\relax IEEE, 2022, pp. 631--636.

\bibitem{9530497}
G.~Shi, Y.~Xiao \emph{et~al.}, ``{From Semantic Communication to Semantic-Aware Networking: Model, Architecture, and Open Problems},'' \emph{{IEEE Communications Magazine}}, vol.~59, no.~8, pp. 44--50, 2021.

\bibitem{10454584}
S.~E. Trevlakis, N.~Pappas \emph{et~al.}, ``{Toward Natively Intelligent Semantic Communications and Networking},'' \emph{IEEE Open Journal of the Communications Society}, vol.~5, pp. 1486--1503, 2024.

\bibitem{9679803}
X.~Luo, H.-H. Chen \emph{et~al.}, ``{Semantic Communications: Overview, Open Issues, and Future Research Directions},'' \emph{IEEE Wireless Communications}, vol.~29, no.~1, pp. 210--219, 2022.

\bibitem{Wheeler2022Engineering}
D.~Wheeler and B.~Natarajan, ``{Engineering Semantic Communication: A Survey},'' \emph{IEEE Access}, vol.~11, pp. 13\,965--13\,995, 2022.

\bibitem{9979702}
W.~Yang, Z.~Q. Liew \emph{et~al.}, ``{Semantic Communication Meets Edge Intelligence},'' \emph{IEEE Wireless Communications}, vol.~29, no.~5, pp. 28--35, 2022.

\bibitem{lu2023semantics}
Z.~Lu, R.~Li \emph{et~al.}, ``{Semantics-Empowered Communications: A Tutorial-Cum-Survey},'' \emph{IEEE Communications Surveys \& Tutorials}, vol.~26, no.~1, pp. 41--79, 2024.

\bibitem{Iyer2022A}
S.~Iyer, R.~Khanai \emph{et~al.}, ``{A Survey on Semantic Communications for Intelligent Wireless Networks},'' \emph{Wireless Personal Communications}, vol. 129, pp. 569--611, 2022.

\bibitem{li2023open}
P.~Li and A.~Aijaz, ``{Open RAN Meets Semantic Communications: A Synergistic Match for Open, Intelligent, and Knowledge-Driven 6G},'' \emph{arXiv preprint arXiv:2310.09951}, 2023.

\bibitem{strinati2024goal}
E.~C. Strinati, P.~Di~Lorenzo \emph{et~al.}, ``{Goal-Oriented and Semantic Communication in 6G AI-Native Networks: The 6G-GOALS Approach},'' \emph{arXiv preprint arXiv:2402.07573}, 2024.

\bibitem{getu2024performance}
T.~M. Getu, W.~Saad \emph{et~al.}, ``{Performance Limits of a Deep Learning-Enabled Text Semantic Communication under Interference},'' \emph{IEEE Transactions on Wireless Communications}, pp. 1--1, 2024.

\bibitem{shokrnezhad2024acnc}
M.~Shokrnezhad, H.~Yu \emph{et~al.}, ``{Towards a Dynamic Future with Adaptable Computing and Network Convergence (ACNC)},'' \emph{arXiv preprint arXiv:2403.07573}, 2024.

\bibitem{our_cl_d3ql_paper}
H.~Mazandarani, M.~Shokrnezhad, and T.~Taleb, ``{A Semantic-Aware Multiple Access Scheme for Distributed, Dynamic 6G-Based Applications},'' \emph{arXiv preprint arXiv:2401.06308}, 2024.

\bibitem{getu2023making}
T.~M. Getu, G.~Kaddoum \emph{et~al.}, ``{Making Sense of Meaning: A Survey on Metrics for Semantic and Goal-Oriented Communication},'' \emph{IEEE Access}, vol.~11, pp. 45\,456--45\,492, 2023.

\end{thebibliography}




\end{document}